\documentclass{iauc}
\usepackage{graphicx}
\pubyear{2005}
\volume{199}
\pagerange{1--}
\jname{Probing Galaxies through Quasar Absorption Lines}
\editors{P. R. Williams, C. Shu, and B. M\'{e}nard, eds.}

\title[The Metallicity - Kinematics Relation in Large-N(HI) Absorbers] 
{The Mean Metallicity - Kinematics Relation in High Column Density MgII Absorbers
and Selection Effects in DLA Surveys}

\author[Turnshek, Rao, Nestor, Belfort-Mihalyi, Quider]   
{David A. Turnshek$^{1,2}$, Sandhya M. Rao$^1$, Daniel B. Nestor$^{1,3}$, \break
Mich\`ele Belfort-Mihalyi$^1$, \and Anna M. Quider$^1$}

\affiliation{$^1$Department of Physics and Astronomy, University of Pittsburgh,
Pittsburgh, PA 15260, USA \break 
$^2$Email: turnshek@pitt.edu, $^3$Present address: Department of Astronomy, University of Florida}

\begin{document}

\maketitle

\begin{abstract}
Sloan Digital Sky Survey (SDSS) quasar spectroscopy is 
yielding a database of strong low-ionization MgII absorbers over the redshift
interval $0.36<z<2.28$ which is over two orders of magnitude larger 
than anything previously assembled. {\it Hubble Space Telescope} (HST) UV spectroscopy has been 
used to measure neutral hydrogen column densities for a small subset 
of them. These data empirically show that MgII absorbers 
with rest equivalent widths $W^{\lambda2796}_0 \ge 0.6$ \AA\ have 
a mean neutral hydrogen column density that is roughly constant at 
$N(HI) \approx 4\times10^{20}$ atoms cm$^{-2}$, with individual systems
lying in the damped Ly$\alpha$ (DLA) and sub-DLA regimes. Since the MgII doublets
generally exhibit saturation, the $W^{\lambda2796}_0$ values are an indication 
of the absorbers' velocity spreads. Thus, we can study neutral-gas-phase
metallicities as a function of kinematics by forming SDSS composite spectra 
and measuring weak unsaturated metal lines that form in neutral gas (e.g., CrII, FeII, MnII, SiII, ZnII)
as a function of $W^{\lambda2796}_0$. We use this method on SDSS composite spectra
to show how metallicity and kinematics are positively correlated for large-$N(HI)$ 
absorbers, including trends related to dust depletion and
the enhancement of $\alpha$-elements. We also discuss the need to account for 
selection effects in DLA surveys, and we make inferences about models for 
DLA absorption and their contribution to cosmic star formation.


\end{abstract}

\firstsection 

\section{The Properties of High Column Density Absorbers}

Rao (2005; {\it this volume}) has reviewed the statistics of DLAs, with some 
emphasis on results at redshifts $z<1.65$ based on HST UV spectroscopy. See also
Rao, Turnshek, \& Nestor (2005a, hereafter RTN2005). It is widely believed that the 
classical DLAs with $N(HI) \ge 2\times10^{20}$ atoms cm$^{-2}$ track the 
neutral gas component of the universe at $z<4$ (but see \S4).  The measurements indicate that the 
neutral gas mass in DLAs is approximately constant from $z\approx4$ to 
$z\approx0.5$, but drops by a factor of $\approx2$ from $z\approx0.5$ to $z=0$.
Also, relative to no-evolution, the product of gas cross section and comoving number density of absorbers 
drops by a factor of $\approx2$ from $z\approx4$ to $z\approx2$, but then follows the 
no-evolution curve from $z\approx2$ to $z=0$ ($h=0.7$, $\Omega=0.3$, $\Lambda=0.7$).
At $z<1$, where searches for galaxies associated with DLAs have 
been reasonably successful, a mix of galaxy types have been found, e.g., dwarfs, low surface brightness (LSB) 
galaxies, luminous spirals, etc. (see Rao et al. 2003 and references therein). 
In addition, the mean column-density-weighted 
metallicity of the DLAs increases from [X/H] $\approx-1.6$ at $z\approx4$ to 
[X/H] $\approx-0.8$ at $z\approx1$ where X is Zn, Fe, O, or Si (e.g., see Rao et al. 2005b
and references therein). Aside from these trends, 
we also previously reported a positive correlation between mean metallicity and kinematics in high-$N(HI)$ 
systems (Nestor et al. 2003); one of the aims of this contribution is to discuss 
more recent work on the details of this relationship (\S2).
We emphasize that while substantial observational progress on DLAs has been made over the last five years
({\it this volume}),
properly connecting these results to scenarios for galaxy and structure formation will 
require careful consideration of selection effects; we will consider several applications 
to illustrate this (\S3). We will also discuss the implications of our results and other recent work (\S4)

\section{The Metallicity - Kinematics Relation in Large-N(HI) Absorbers}

\subsection{Formation of SDSS Composite Spectra vs. MgII Rest Equivalent Width}

The MgII$\lambda\lambda2796,2803$ absorbers visible in SDSS quasar spectra cover 
the redshift interval $0.36<z<2.28$. The number of systems in the completed SDSS will be well 
over two orders of magnitude larger than anything previously assembled. 
As is now well established, the DLAs are traced by a subset of the strong MgII absorbers,
and track the bulk of the HI gas mass that has so far been observed in
the universe. In the study described here we are interested in using the MgII absorbers to investigate 
the mean relationship between metallicity and velocity spread in large-$N(HI)$ 
absorbers.  The idea is as follows. HST UV spectroscopy has been 
used to measure HI column densities for a small subset of the SDSS MgII systems, each
of which has a known rest equivalent width (REW), $W^{\lambda2796}_0$ (Rao 2005, 
{\it this volume}; RTN2005). From these data we know empirically that for MgII 
absorption lines with $0.3 \le W^{\lambda2796}_0 < 0.6$ \AA\ the mean HI column density is
$N(HI) \approx 10^{19}$ atoms cm$^{-2}$; for $W^{\lambda2796}_0 \ge 0.6$ \AA\ it  
is roughly constant at $N(HI) \approx 4\times10^{20}$ atoms cm$^{-2}$ with
individual $N(HI)$ values spanning about 3 orders of magnitude and falling within 
both the sub-DLA and DLA regimes (typically $10^{18.7} < N(HI) < 10^{21.7}$ atoms cm$^{-2}$). 
However, the MgII doublets generally exhibit saturation, so the $W^{\lambda2796}_0$ 
values are an indication of the absorbers' velocity spreads. Thus, we can study 
neutral-gas-phase metallicities as a function of kinematics by forming SDSS composite 
spectra and measuring weak unsaturated metal lines as a function of $W^{\lambda2796}_0$. 
To accomplish this we followed a method similar to the one described by Nestor et al. 
(2003), who studied the neutral gas phase element abundances of Zn and Cr. The MgII 
absorbers incorporated in this analysis were identified in SDSS quasar spectra 
up to DR3 using the Nestor, Turnshek, \& Rao (2005a) selection criteria. Nearly 6000 
absorption systems were used to form composite spectra for different intervals of  
$W^{\lambda2796}_0$ in regions covering MgII and MgI, and 
the low oscillator strength metal lines due to MnII, 
FeII, ZnII, CrII, and SiII. These composites approximately covered the redshift 
interval $1<z<2$, as dictated by the rest wavelengths of the metal lines 
and the short wavelength cutoff of SDSS at $\approx 3800$ \AA.
As examples, the left side of Figure 1 shows the composites covering the ZnII and CrII region,
while the right side shows the composites for the MgII and MgI region.

\begin{figure}
\centerline{\scalebox{0.55}{\rotatebox{270}{\includegraphics{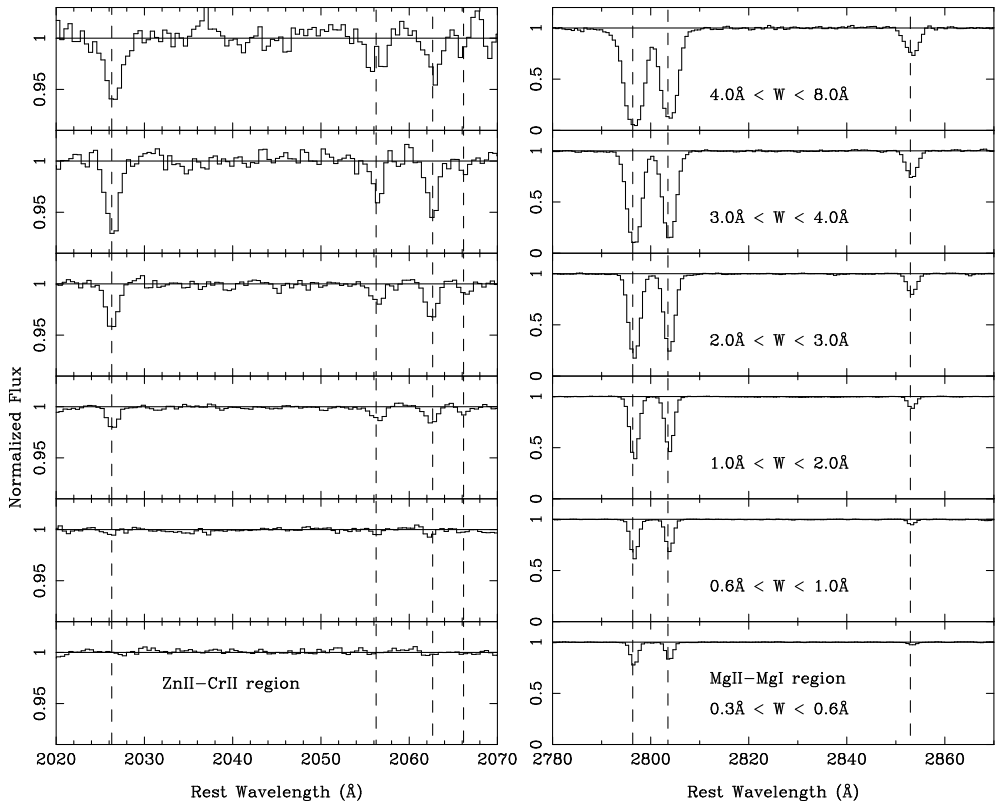}}}}
\caption{The left side shows the flux-normalized ZnII and CrII 
region of the composite spectra, while the right side shows the 
corresponding MgII and MgI region. The $N(HI)$ value is constant 
at $\approx4\times10^{20}$ atoms cm$^{-2}$ in all of the composite spectra except
for the bottom most panels where $N(HI)\approx10^{19}$ atoms cm$^{-2}$.
$W^{\lambda2796}_0$ increases from bottom to top. The fact that the MgII doublet is
saturated indicates that increases in $W^{\lambda2796}_0$ are primarily due to larger
velocity spreads in the metal-line absorption. At the same time, the much weaker 
REWs of the ZnII and CrII metal lines (note the scale on the normalized flux) indicate 
that these lines are unsaturated, hence the increase in their REWs from bottom to
top is due to an increase in metallicity.}
\end{figure}

\subsection{Correlation between Mean Metallicity and Velocity Spread}

Measurements of weak metal lines due to MnII, FeII, ZnII, CrII, and SiII
in the composite spectra resulted in the mean element abundance determinations 
graphically illustrated in Figure 2. 
In converting the measured REWs to metal-line column densities, saturation 
effects in the weak lines were assumed to be unimportant. This is 
consistent with the finding that our results are  robust and independent of 
inclusion of the handful of cases where these metal lines are seen in individual 
SDSS spectra. In addition, possible ionization corrections were not made;
thus the results incorporate the assumption that singly ionized states of Mn, Fe, Zn, Cr, and Si 
are present in the neutral regions that contribute to the 
measured mean HI column density of $N(HI) \approx 4\times10^{20}$ atoms
cm$^{-2}$ at $W^{\lambda2796}_0 \ge 0.6$ \AA. 
For Figure 2 we transformed $W^{\lambda2796}_0$ in units of
\AA\ (see Figure 1) to $\Delta$V$_{REW} = c(W^{\lambda2796}_0/2796)$ in units of km s$^{-1}$, since 
this more explicitly makes the point that the $\lambda2796$ line is saturated
and that $W^{\lambda2796}_0$ is most appropriately indicative of a velocity 
width or spread (although $\Delta$V$_{REW}$ is not the Doppler parameter and $W^{\lambda2796}_0$
does increase as the log of the product of column density and oscillator
strength in this regime, so these velocity values need to be interpreted 
carefully). 

We see that the overall
level of metal enrichment in strong MgII absorption systems can
be separated out simply on the basis of $W^{\lambda2796}_0 \equiv
\Delta$V$_{REW}$, i.e., a positive relationship between mean metallicity and
kinematics. However, within this overall relationship, several notable 
trends are present. We discuss these separately below.

{\bf (1) Fe, Zn, and Cr.} These elements are often assumed to track 
the Fe-peak elements and to have a common origin.
Results indicate that Zn suffers little depletion in DLA absorbers, so
throughout we will assume that Zn is not depleted, although some Zn depletion
might be expected at the highest metallicities.
Measurements of the FeII, ZnII, and CrII lines at the lowest velocity spreads 
are uniformly consistent with a mean abundance of [X/H] $\approx -2.0$ (1\% solar) and 
therefore Fe and Cr show no evidence for depletion. However, the ZnII measurement at the highest velocity spreads
yields a mean abundance of [Zn/H] $\approx -0.4$ (40\% solar), and the more depleted elements of Fe and 
Cr have abundances of [Fe/H] $\approx -1.1$ (i.e., [Fe/Zn] $\approx -0.7$, or Fe is at least $\approx80$\% depleted)
and [Cr/H] $\approx -0.95$ (i.e., [Cr/Zn] $\approx -0.55$, or Cr is at least $\approx72$\% depleted). 
The mean metallicity - kinematics relation for Zn in strong MgII absorbers can be parameterized
as [Zn/H] $=-0.4-0.0043e^{[(6-W^{\lambda2796}_0)/0.88]}$, where 
$0.6$ \AA\ $\le W^{\lambda2796}_0 < 6.0$ \AA.
Thus, there is a clear trend showing
the overall increase in mean metallicity with increasing velocity spread ($W^{\lambda2796}_0$);
and depletion increases with increasing mean metallicity,
although it should be pointed out that the nucleosynthetic
processes which lead to the formation of Zn are still debated (e.g.,
Umeda \& Nomoto 2002).

\begin{figure}
\centerline{\scalebox{0.77}{\rotatebox{270}{\includegraphics{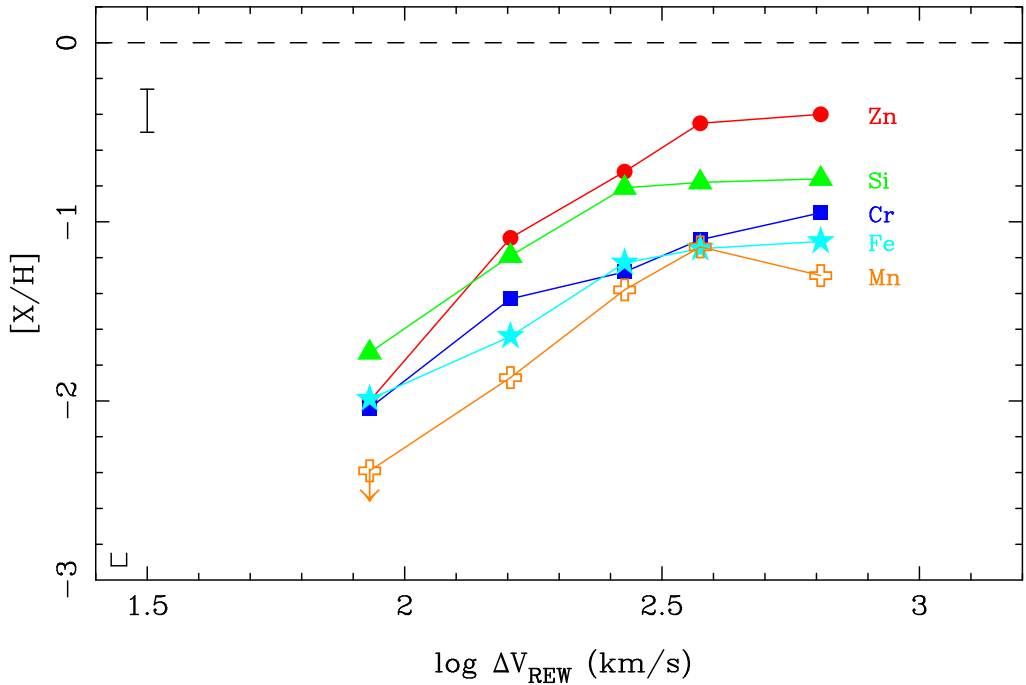}}}}
\caption{The metallicity of Zn, Si, Cr, Fe, and Mn as a function of $W^{\lambda2796}_0$
in velocity units, $\Delta$V$_{REW}$, which is roughly correlated with the 
velocity spread of the absorption. As discussed in the text, at small REW velocity
the metallicity is low, there is little evidence for depletion on to grains, the Mn 
abundance is significantly lower than the other Fe-peak elements,
and the $\alpha$-element Si is enhanced. 
At large REW velocity there is significant evidence for
depletion on to grains. The redshift interval covered by these measurements 
generally lie in the range $1<z<2$.}
\end{figure}

{\bf (2) Mn.} The Fe-peak element Mn displays a somewhat different behavior.
We note that the work discussed by Savage \& Sembach (1996) indicates that Mn and
Fe are similarly depleted in warm Galactic halo clouds ([Mn/Fe] $\approx 0$), but 
that Fe is significantly more depleted than Mn in cool Galactic disk clouds ([Mn/Fe] $\approx +0.8$).
Our results for the strong MgII systems 
show that [Mn/Fe] $\approx -0.1$ at the highest velocity spreads when [Zn/H] $\approx -0.4$ 
and that [Mn/Fe] $< -0.4$ at the lowest velocity spreads when [Zn/H] $\approx -2$.
This trend of lower relative Mn abundance with decreasing metallicity 
is consistent with individual DLA measurements reported in figure 6 of 
Dessauges-Zavadsky, Prochaska, \& D'Odorico (2002), and is also roughly
consistent with Mn measurements for Galactic stars (no depletion).
Mn is produced mostly by SNeIa, so a common qualitative interpretation of the low [Mn/Fe] values
in low-metallicity DLAs is that the gas is in a relatively early stage of chemical evolution,
and thus has only undergone significant enrichment by SNeII (e.g., Lu et al. 1996). 

{\bf (3) Si.} The $\alpha$-element 
Si shows interesting trends relative to (assumed) undepleted Zn.
At the highest velocity spreads when [Zn/H] $\approx -0.4$ we find  
[Si/Zn] $\approx -0.35$, which is likely representative of some Si depletion, but empirically Si is 
likely less depleted than Fe (Savage \& Sembach 1996), and this is consistent with our measured value 
of [Fe/Zn] $\approx -0.7$.
At the lowest velocity spreads when [Zn/H] $\approx -2$ we find [Si/Zn] $\approx +0.25$ and [Fe/Zn] $\approx 0$.
Thus, our results indicate that the $\alpha$-element Si is 
enhanced at these low metallicities.
This is generally consistent with the well established result that very 
metal-poor Galactic halo stars show enhanced levels of $\alpha$-elements (e.g., McWilliam 1997).
Once again, the results are qualitatively explained by having the low metallicity 
gas being in a relatively early stage of chemical evolution, without significant 
enrichment by SNeIa. However, the enhancement of Si in metal poor Galactic stars 
(e.g., McWilliam 1997) appears to be somewhat more extreme than what we find for our strong MgII sample.

In general, altering the details of the initial mass function (IMF), the star formation rate (SFR),
and mixing mechanisms
will affect the abundance trends present in gas with cosmic time. For example, in the case 
described above one might infer that Si at $1<z<2$ is less enhanced in our sample at low metallicity 
(in comparison to similarly low metallicities in present-day Galactic stars) because of comparatively lower
SFRs. Indeed, Si enhancement at low metallicity in our Galaxy is smaller 
at larger galactocentric distance for fixed metallicity, and this can be 
attributed to a lower SFR at larger galactocentric distance (Edvardsson et al. 1993).

An interesting exercise is to compare the mean metallicity - kinematics relation 
derived from composite spectra (Figure 2) to individual 
measurements of DLA metallicities from high-resolution spectroscopy (Figure 3). 
The individual metallicities shown in Figure 3 correspond to redshifts $1.78 < z < 4.22$, while
the mean metallicities shown in Figure 2 correspond to $1<z<2$. Of course, comparison of the two figures is
complicated not only by the redshift difference, but also by the fact that one result incorporates all 
strong MgII absorbers while the other is limited to DLAs. 
The metallicities used in Figure 3 are generally 
taken from tabulations by Prochaska and collaborators (e.g., references given in 
Prochaska et al. 2003), but the $\Delta$V values 
are representative of the velocity extent over which low-ionization absorption is clearly visible
(not the velocity interval containing 90\% of the gas, which was kindly provided to us for
comparison by Jason Prochaska). In either case
we find that individual DLA measurements show a trend of increasing metallicity with increasing
velocity spread, but the slope becomes flatter when the velocity interval containing 90\% of
the gas is used. This trend is also seen in the high-resolution ESO VLT-UVES data presented 
by Ledoux et al. (2005) at this symposium. 

The results for strong MgII absorbers described above represent
gas-phase abundance trends averaged over all neutral hydrogen clouds in all types of absorbing galaxies
for a large range of inclination angles and galactocentric distances as sampled 
by SDSS quasar sightlines. We suspect that the lower metallicity sightlines will preferentially correspond to 
the progenitors of present-day dwarfs, LSB galaxies, and the outer regions 
of luminous galaxies, while the higher metallicity sightlines will primarily 
sample the more chemically evolved regions of more luminous galaxies, often at lower impact parameters. 
Thus, the observed relationship between mean metallicity and kinematics provides an empirical
constraint on the kinematic and/or dynamic evolution of gaseous regions in galaxies (e.g., disk,
halo, and protogalactic merging components). 
It is natural to think that the more chemically evolved gaseous regions also acquire more mechanical
energy (e.g., from gravitational processes and SN explosions) which results in these regions exhibiting
greater velocity spreads on average.

\begin{figure}
\centerline{\scalebox{0.75}{\rotatebox{0}{\includegraphics{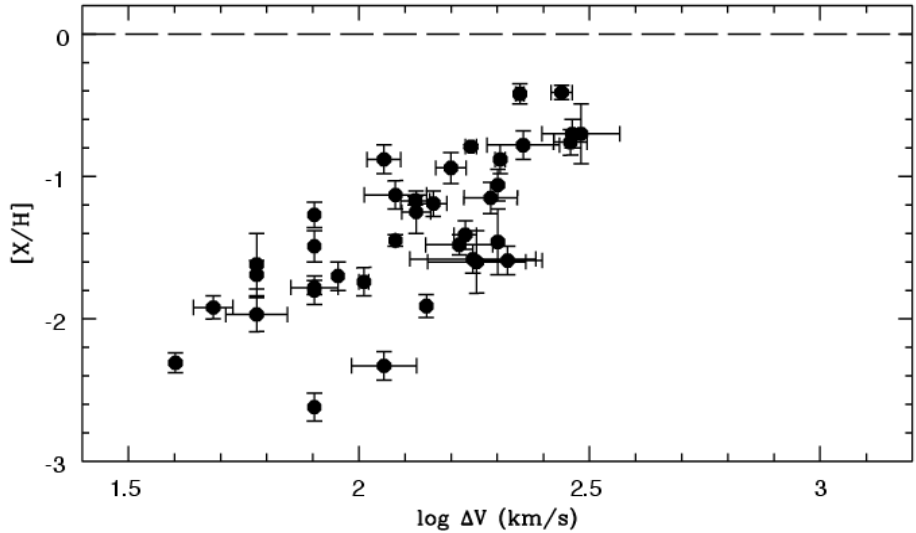}}}}
\caption{The metallicity - kinematics relation inferred from a subset of DLAs 
for which high-resolution Keck spectroscopy is readily available (e.g., references
given in Prochaska et al. 2003). These DLA absorbers are generally at higher redshift, 
$1.78<z<4.22$, in comparison to the redshift range covered in the SDSS composite spectra.
The $\Delta$V parameters, estimated visually from published high-resolution 
spectra, are the velocity intervals over which low-ionization 
metal-line absorption is clearly present. There is a positive
correlation between metallicity and $\Delta$V, consistent with that seen in 
the SDSS composite spectra.}
\end{figure}

\section{Selection Effects in Galaxy Surveys}

\subsection{Gas-Cross-Section Selection vs. Brightness/Color Selection}

The most common type of galaxy surveys are based on multi-color imaging. Since these
are magnitude-limited surveys, objects such as dwarf and LSB 
galaxies have a tendency to be missed, especially with increasing redshift. 
Alternatively, quasar absorption-line observations can be used to conduct 
completely different types of galaxy surveys. Those which rely on the MgII 
doublet and/or DLA line identifications are galaxy surveys based on gas-cross-section 
selection. Yet surveys which use strong MgII selection have a 
very different physical basis than those which use blind DLA selection.

If reddening/dimming due to dust and amplification due to gravitational lensing
were negligible in all DLA systems (but see M\'enard et al. 2005), a blind 
DLA survey yielding a compilation of results as a function of $N(HI)$ would 
be the most straight forward way to perform a survey for galaxies based on HI gas 
cross section. However, the incidence of DLAs is small enough, 
especially at low redshift, that we are forced to use the absence of  
low-ionization metal lines (e.g., the otherwise easily-detected MgII doublet) as a means to 
eliminate sightlines which would have been observed with negative results in any 
blind DLA survey. As discussed in \S2, $W^{\lambda2796}_0$ is the most important indicator,
since there is little chance of encountering a DLA unless $W^{\lambda2796}_0 \ge 0.6$ \AA\
(Rao \& Turnshek 2000; RTN2005; Rao 2005, {\it this volume}).
This threshold value, $W^{\lambda2796}_0 \equiv \Delta$V$_{REW} \ge 65$ km s$^{-1}$ in velocity units, 
is the physical basis for MgII-selected surveys and it indicates that DLAs 
reside in environments that have some minimum value for the velocity spread. Fortunately,
high to moderate quality SDSS quasar spectroscopy permits the detection of 
MgII doublets at this threshold. 

At the same time, it is important to recognize that the properties of a 
MgII-selected sample (e.g., the incidence of DLAs in such a sample) will be biased 
unless the observed sample's $W^{\lambda2796}_0$ 
distribution matches the true $W^{\lambda2796}_0$ distribution. The
true $W^{\lambda2796}_0$ distribution
has recently been determined using SDSS data (Nestor et al. 2005a; figures 1 and 2 from
Nestor, Turnshek, \& Rao 2005b, {\it this volume}), so it is now possible to assess any
mis-match between the observed and true distribution and correct for it if necessary. 
Below we describe how MgII selection may bias the properties of a sample.

\subsection{Examples: Incidence, $N(HI)$ Distribution, Metallicities, DLA Galaxies}

The criteria adopted in a MgII-selected DLA survey 
may affect the properties of the resulting DLA absorber sample in terms 
of: (1) the incidence of DLAs per unit redshift, 
(2) the $N(HI)$ column density distribution of identified DLAs, 
(3) the resulting DLA metallicities, and most likely (4) the properties of identified DLA galaxies.
We briefly discuss each of these effects below.

{\bf (1) Dependence of the DLA incidence on $W^{\lambda2796}_0$.}
As seen in figure 4 of Rao (2005, {\it this volume}; 
also RTN2005), the fraction of MgII absorbers
which are DLAs with $N(HI) \ge 2\times10^{20}$ atoms cm$^{-2}$ 
depends on $W^{\lambda2796}_0$, rising from a fraction near 16\%
just above the threshold value of $W^{\lambda2796}_0=0.6$ \AA\ to 
about 65\% at the highest values near $W^{\lambda2796}_0=3$ \AA.
At present these should be considered approximate fractions,
since the REW of FeII$\lambda2600$ can be used (and has been used)
to increase the probability of discovering a DLA, however this is a
good assessment of the fractions to first order. Thus, unless the observed
sample's $W^{\lambda2796}_0$ distribution matches the true distribution, a bias
will be introduced into the determination of the DLA incidence.
The methods used in RTN2005 account for these differences when calculating the DLA
incidence.

{\bf (2) Possible dependence of the DLA N(HI) distribution on $W^{\lambda2796}_0$.}
The degree to which $N(HI)$ may be biased by $W^{\lambda2796}_0$ can be seen 
by examining the right panel in figure 4 of Rao (2005, {\it this volume}; also
RTN2005). The mean HI column density of identified
DLAs is $N(HI) \approx 2.5\times10^{21}$ atoms cm$^{-2}$ when 
$0.6$ \AA\ $\le W^{\lambda2796}_0 < 1.2$ \AA, but it seems to decrease by a
factor of $\approx4$ at $W^{\lambda2796}_0\approx3$ \AA.
However, inspection of figure 2 of Rao (2005, {\it this volume}) suggests
that this trend is not particularly tight nor is it well established for DLAs
by themselves. On the other hand, if one considers all the points in
figure 2 of Rao (2005), it is clear that the $N(HI)$ distribution changes for 
different $W^{\lambda2796}_0$ intervals.
It is interesting that in MgII-selected surveys for DLAs, 
the determination of the cosmological mass density of neutral 
gas, $\Omega_{DLA}$, has (so far) not revealed any dependency 
on $W^{\lambda2796}_0$ selection. This is because
the increased probability (by a factor of $\approx 4$) of finding a DLA 
at the largest $W^{\lambda2796}_0$ values is approximately compensated for by the 
corresponding decrease in mean HI column density (by a factor
of $\approx 4$) at the largest $W^{\lambda2796}_0$ values. 
It is worth pointing out that although the MgII selection criteria which have been 
employed in low-redshift DLA surveys lead to reasonably complete
samples of DLAs, incompleteness must set in at HI column densities in the 
sub-DLA regime, because systems with $W^{\lambda2796}_0 <0.3$ \AA\ can have 
sub-DLA HI column densities. Therefore, only the $N(HI)$ distribution in the DLA regime 
can be reliably considered with available data. 

{\bf (3) Dependence of mean metallicity on $W^{\lambda2796}_0$.}
The analysis described in \S2 demonstrated that the mean 
metallicity of a MgII-selected sample is a strong function of 
$W^{\lambda2796}_0$. Although the mean HI column densities of the
binned samples were in the DLA regime and constant at $N(HI) \approx 4\times10^{20}$
atoms cm$^{-2}$, most of the individual
absorbers are not DLAs. Indeed, above we
noted that only $\approx16$\% of the sample are DLAs 
just above the threshold value of $W^{\lambda2796}_0=0.6$ \AA,
and $\approx65$\% are DLAs near $W^{\lambda2796}_0=3$ \AA.
Therefore, the results of our metallicity determinations in \S2 apply to strong
MgII absorbers, not just DLAs. 
Of course, it should be kept in mind that in a fixed $W^{\lambda2796}_0$ interval 
we are sampling a range of $N(HI)$ values, and the sampled systems undoubtedly 
exhibit a range of metallicities. 
Thus, a significant amount of detail remains to be discovered through future work. 
In the end, a reliable determination of the statistical details of cosmic metallicities will
depend on unbiased sampling of the $W^{\lambda2796}_0$ distribution,
or making a correction for biased sampling. The extent to which
such biases may have affected existing results 
(e.g., Prochaska et al. 2003, Rao et al. 2005b) is unclear.

{\bf (4) Possible dependence of DLA galaxy type on $W^{\lambda2796}_0$.}
Finally, given the various biases discussed above, it is of interest to 
consider how an observed sample's $W^{\lambda2796}_0$ distribution may bias the results
of searches for DLA galaxies. Significant statistics 
have not yet been compiled on this issue. Inspection of the results on
DLA galaxy identifications (Rao et al. 2003) shows no strong
correlations with $W^{\lambda2796}_0$, but sample sizes are still small.
Nestor et al. (2005b; {\it this volume})
have presented some preliminary evidence which hints at the possibility that 
very luminous galaxies preferentially lie along the sightlines to ultra-strong MgII absorbers.
In any case, the issue of how MgII selection methods bias the type, luminosity, and 
impact parameter of identified absorbing galaxies needs to be investigated. Owing
to the correlation between mean metallicity and velocity spread (\S2.2), some bias is expected.
Thus, this inadvertent bias should be accounted for when evaluating, for example, 
the true frequency of galaxy morphologies, luminosities, and impact parameters 
that give rise to DLA absorbers.

\section{Implications}
We have presented the results of an analysis which reveals a positive correlation between mean metallicity and 
gaseous velocity spread in strong MgII absorbers. We have also speculated that this
relationship is predominantly due to the low-metallicity, low-velocity-spread
sightlines more preferentially sampling the progenitors of dwarfs, LSB galaxies, and 
the outer regions of luminous galaxies (primarily because sightlines through such regions
are expected to have low metallicity); whereas the very rare higher-metallicity, higher-velocity-spread 
sightlines should sample the more chemically evolved regions associated with more 
luminous galaxies. Rare sightlines which intercept 
Lyman break galaxies (LBGs), which can exhibit high metallicity and large velocity
spreads in the low-ionization ISM lines seen in their spectra (e.g., Shapley et al. 2003),
may be examples of this latter case. SNeII explosions may drive large bubbles of hot gas out of LBGs
and entrain neutral clouds which give rise to DLAs in the flow. 
But the $W^{\lambda2796}_0$ distribution and the fraction of DLAs as a function of $W^{\lambda2796}_0$
indicates that such situations are rare, happening $<20$\% of the time, and therefore are not usually 
responsible for the DLAs.

An important empirical result which needs to be understood, and which is perhaps more 
fundamental to explaining the bulk of the DLAs, is 
why $W^{\lambda2796}_0 \equiv \Delta$V$_{REW} \ge 65$ km s$^{-1}$ is required to find a DLA. 
The model put forth by 
Mo, Mao, \& White (1998) may provide the beginnings of a scenario to explain this. 
They studied the formation of galactic disks in cold dark matter (CDM) hierarchical models 
for structure formation.
In their model the bulk of DLAs are produced by centrifugally supported disks within
dark matter (DM) halos. The large angular momentum would make
these disks more extended and lower in surface brightness (i.e., lower SFRs) in comparison to a 
low-angular-momentum spheroid that would be more compact and have higher surface brightness (i.e., higher SFRs). 
Mo, Mao, \& White (1999) suggest that such spheroids may correspond to LBGs. 
Mo et al. (1998) calculate expected impact parameters for DLA
disks for various circular velocities that might be required for disk stability.
Interestingly, their model predictions are consistent 
with observed impact parameters of low-redshift DLA 
galaxies (e.g., compare figure 12 in Mo et al. 1998 with 
figure 10 in Rao et al. 2003) when circular velocities of 
$v_{cir} > 100$ km s$^{-1}$ is taken as a requirement for stable disks.
Although the details of such a model may need to be 
modified, one can envision that this type of effect, typically coupled with a reduction in sightline
velocity spread due to inclination effects, may largely be responsible for the fact that 
$W^{\lambda2796}_0 \equiv \Delta V_{REW} \ge 65$ km s$^{-1}$ is required for a sightline 
to encounter a DLA. Mo et al. (1998, 1999) comment that various other mechanisms may
also be important, and this may depend on the redshift regime.
Merging systems of protogalactic clumps (Haehnelt, Steinmetz, \& Rauch 1998) 
and non-equilibrium disks in low-mass DM halos may also contribute to the DLAs. As noted above,
LBGs may also from time to time be associated with DLAs, but at much lower incidence since 
their low angular momentum makes them much less extended. 
Within all of these possibilities there may not be 
significant total gas cross section to generally give rise to a DLA component
below the line-of-sight threshold of 
$W^{\lambda2796}_0 \equiv \Delta V_{REW} \approx 65$ km s$^{-1}$. This is an important
constraint on future CDM simulations that seek to explain DLAs.

The possibility that some types of regions in the universe may have too 
small an interception probability (i.e., gas cross section times comoving number density
of that type of region) to be easily identified as DLAs in MgII or blind DLA 
searches is an important issue for another reason. 
Hopkins, Rao, \& Turnshek (2005; also Rao 2005, {\it this volume}) have 
discussed the question of whether the observed population of DLAs 
can account for the observed cosmic SFR from low to high redshift. 
By applying the Kennicutt (1998) formulation of the Schmidt law to 
the properties of the currently observed population of DLAs they find
that the DLAs under-predict the cosmic SFR inferred from the
luminosity density of high-redshift galaxies (see figure 12 in Rao 2005, {\it this 
volume}). An even larger discrepancy occurs when one compares the DLA 
metallicities to the metallicities expected on the basis of the cosmic SFR (see
figure 13 in Rao 2005, {\it this volume}). One way to avoid this discrepancy 
is to postulate that the MgII and blind DLA surveys
are not yet large enough to include absorbers with 
very small individual cross sections that nevertheless may
dominate the cosmic SFR and be the main reservoirs for the metals. 

Indeed, these star forming regions will be rich in molecular clouds,
even though HI column densities may still exceed H$_2$ column densities
along sightlines passing through them.
Of course molecular clouds, not HI clouds, most directly provide
the fuel for star formation. Kennicutt (1998) points out that in normal
disks star formation generally takes place in regions that contain
$1-100$ M$_{\odot}$ pc$^{-2}$ (i.e., $\approx 10^{20} - 10^{22}$ atoms or molecules cm$^{-2}$),
whereas the more rare (and smaller) star burst regions contain
$10^2 - 10^5$ M$_{\odot}$ pc$^{-2}$ (i.e., $\approx 10^{22} - 10^{25}$ atoms or molecules cm$^{-2}$).
It therefore seems reasonable to conclude that most of the neutral and molecular gas mass has so far been missed in DLA surveys,
and this seems consistent with the fact that molecules are rarely seen in identified DLA absorbers
(e.g., Ledoux, Petitjean, \& Srianand 2003). 

In this regard it is interesting that Gardner et al. (1997) 
found in their CDM simulations that depletion of the gas supply by star formation only affected the DLA statistics
at $z>2$ for $N(HI) > 10^{22}$ atoms cm$^{-2}$ (i.e., in a regime where DLAs have not generally been
found), even though roughly half of the cold collapsed gas was
converted to stars by $z=2$.
But if significant mass has been missed due to small total cross section for star forming regions, 
whether or not these high-gas-mass regions will be found 
once sample sizes are much larger is unclear. Substantial dust-induced reddening may
prevent complete samples from ever being discovered via optical quasar absorption-line spectroscopy. 
However, in the end, analysis of intervening absorption-line systems in SDSS quasars
can still provide powerful constraints on the evolution of metals and the properties of
gaseous structures in the universe.

\begin{acknowledgments}
We thank members of the SDSS collaboration who made the SDSS project
a success. We thank Nick Allen and Daniel Owen who helped measure
MgII absorption lines during their participation in an REU program
at the University of Pittsburgh. We also thank Andrew Hopkins for collaborative
work and discussions about cosmic SFRs, Jason Prochaska for providing
information on DLA velocity spreads for individual DLAs, and
Regina Schulte-Ladbeck for discussions about dwarf galaxies.
We acknowledge and are grateful for concurrent collaborative
work, especially with Brice M\'enard, Dan Vanden Berk, and Don York on SDSS MgII
absorption spectra. We acknowledge financial support from NASA-STScI, NASA-LTSA,
and NSF. HST-UV spectroscopy made the $N(HI)$ determinations
possible, while the SDSS spectra made the metallicity measurements possible.  
Funding for creation and distribution of the SDSS Archive has
been provided by the Alfred P. Sloan Foundation, Participating
Institutions, NASA, NSF, DOE, the Japanese Monbukagakusho, and the
Max Planck Society. The SDSS Web site is www.sdss.org. The
SDSS is managed by the Astrophysical Research Consortium for
the Participating Institutions: University of Chicago, Fermilab,
Institute for Advanced Study, the Japan Participation Group,
Johns Hopkins University, Los Alamos National Laboratory, the
Max-Planck-Institute for Astronomy (MPIA), the Max-Planck-Institute
for Astrophysics (MPA), New Mexico State University, University
of Pittsburgh, Princeton University, the United States Naval
Observatory, and University of Washington.
\end{acknowledgments}

\end{document}